\documentclass[doublecol]{epl2} 

\usepackage{xcolor,graphicx}
\usepackage{amssymb,amsmath,graphicx}
\usepackage[utf8]{inputenc}
\usepackage{comment}
\usepackage{braket}
\usepackage[normalem]{ulem}
\usepackage{hyperref,url}
\usepackage{array}
\usepackage{fontawesome}

\newcommand{\Tr}[1]{\text{Tr} #1}

\title{False Onsager relations}
\shorttitle{False Onsager relations} 

\author{Michele Campisi}
\shortauthor{Michele Campisi}

\institute{                    
   NEST, Istituto Nanoscienze-CNR and Scuola Normale Superiore, I-56127 Pisa, Italy
}

\pacs{xxx}{}

\abstract{
Recent research suggests that when a system has a ``false time reversal violation'' the Onsager reciprocity relations hold despite the presence of a magnetic field. The purpose of this work is to clarify that the  Onsager relations may well be violated in presence of a ``false time reversal violation'': that rather guarantees the validity of distinct relations, which we dub ``false Onsager relations''. We also point out that for quantum systems ``false time reversal violation'' is omnipresent and comment that, per se, this has in general no consequence in regard to the validity of Onsager relations, or the more general non-equilibrium fluctuation relations, in presence of a magnetic field.
Our arguments are illustrated with the Heisenberg model of a magnet in an external magnetic field.}

\begin{document}

\maketitle

\section{Introduction}
One open question in current thermodyamic research is whether a heat engine may achieve Carnot efficiency while delivering finite power \cite{Benenti11PRL106,Allahverdyan13PRL111,Campisi16NATCOMM7,Holubec18PRL121,Bonanca19JSM2019,Miura22PRE105}. In a seminal paper \cite{Benenti11PRL106}, Benenti \emph{et al.} have established that, within the framework of linear response theory, Carnot efficiency could be achieved by a thermoelectric device immersed in a magneic field $\boldsymbol{\mathcal B}$ provided the Seebeck coefficient is not even under the reversal of $\boldsymbol{\mathcal B}$:
\begin{align}
S(\boldsymbol{\mathcal B})\neq S(-\boldsymbol{\mathcal B})\, ,
\end{align}
that is, if the thermopower displays the so called ``Umkehreffekt''. While the latter has been experimentally validated in bismuth crystals \cite{Wolfe63PR129},
still the theoretical understanding of the conditions under which such effect is expected to appear is the subject of active research \cite{Saito11PRB84,Luo20PRR2}. Generally, the origin of its absence should be researched in the presence of symmetries that prevent its appearance \cite{Gabrielli96PRL77,Lebowitz97PRL78,Gabrielli99JSP96}. In particular, it has been suggested   \cite{Bonella14EPL108,Luo20PRR2,Carbone22AP441} that the presence of ``false time reversal violations'' results in the Onsager reciprocity relations to be satisfied notwithstanding the presence of a magnetic field, which would forbid the ``Umkehreffekt''.

Following Robnik and Berry \cite{Robnik86JPA19}, with the expression ``false time reversal violation'' we denote the case where the
 standard textbook ``time reversal'' (namely the transformation that flips velocities and angular momenta, including spins, see Eq. (\ref{eq:timerev}) below) is violated while some other anticanonical (for classical system) or antiunitary (for quantum system) symmetry is obeyed.
All such transformations, which we dub here ``unconventional time reversals'' have the property of inverting time, just like the standard time reversal does, despite the possible presence of a magnetic field \cite{Robnik86JPA19}.
As has been pointed out in recent years, in a broad class of cases ``unconventional time reversal" symmetries exist, and it has been suggesting that Onsager relations would not break in those cases 
 \cite{Bonella14EPL108,Gregorio17SYM9,Coretti18MP116,Coretti20PRE102,Luo20PRR2,Carbone22AP441}.
 
Here we show that i) for any generic quantum system, one could always find at least one ``unconventional time reversal'' symmetry independent of the specific form of the Hamiltonian, ii) the presence of ``unconventional time reversal'' symmetries does not generally imply the validity of Onsager relations. We illustrate that with an example of a quantum system featuring several ``unconventional time reversal'' symmetries where the standard Onsager relations are in fact violated.

We shall also remark that the omnipresence of ``unconventional time reversals'' does not imply the validity of the non-equilibrium fluctuation relations  \cite{Campisi11RMP83} in a magnetic field. We shall shed light onto the fact that, nonethelss, that happens provided the ``unconventional time reversal'' symmetry is one and the same at all times, which often occurs in standard models studied in the literature.

\section{Onsager, Onsager-Casimir, and false Onsager relations}
One of the cornerstones of non-equilibrium thermodynamics are the Onsager relations (ORs) \cite{Onsager31PR37,Onsager31PR38,deGroot84Book}. They dictate that 
the matrix of phenomenological linear response coefficients is composed of symmetric and antisymmetric blocks, depending on the time-reversal parity of the thermodynamic forces and fluxes they connect. Within Kubo's linear response theory \cite{Kubo57aJPSJ12}, such relations are expressed as
\begin{align}
L_{A B}(t) &= \vartheta_A \vartheta_B L_{B A}(t)\, ,
\label{eq:onsager}
\end{align}
where $L_{A B}$ is the relaxation function of the quantity $A$, caused by a perturbation of the quantity $B$. The quantities $A$ and $B$ in Eq. (\ref{eq:onsager}) have definite parity 
$\vartheta_{A(B)}$ under time reversal, namely
\begin{align}
\Theta^\dagger A \Theta = \vartheta_A A,\quad \Theta^\dagger B \Theta = \vartheta_B B\, ,
\end{align}
where $\vartheta_A,\vartheta_B=\pm 1$ and 
\begin{align}
\Theta \boldsymbol{\sigma}_i \Theta^\dagger = - \boldsymbol{\sigma}_i, \quad
 \Theta \mathbf{p}_i \Theta^\dagger = - \mathbf{p}_i, \quad  \Theta \mathbf{q}_i \Theta^\dagger =  \mathbf{q}_i\, ,
 \label{eq:timerev}
\end{align}
is the anti-unitary time reversal operator \cite{Messiah62Book}. 

In Kubo's theory, the relaxation function reads, for a quantum system
\begin{align}
L_{AB}(t) = \int_0^\beta ds \Tr\, \rho A^H(-is) B^H(t) - \Tr \,\rho A B\, ,
\label{eq:relaxation}
\end{align}
where $\rho= e^{-\beta H}/Z$ is the thermal state, and  
$O^H(t)$ denotes the Heisenberg representation of operator $O$ at time $t$, that is
\begin{align}
O^H(t) = U_{t}^\dagger O U_t\, ,
\end{align}
where $U_t =e^{-i H t/\hbar} $
is the unitary time evolution operator and $H$ the system Hamiltonian.
Under the assumption of time-reversal invariance, i.e., 
\begin{align}
 H \Theta =  \Theta H\, ,
\end{align}
we have
\begin{align}
U_{-t} =\Theta U_{t} \Theta^\dagger\, ,
\end{align}
which follows directly from the antiunitary character of the operator $\Theta$. The latter equation says that the application of $\Theta$ causes the inversion of the time evolution. This fact is often refreed to as the principle of microreversibility \cite{Messiah62Book}. Microreversibility combined with the standard rules of quantum mechanics directly implies the validity of the ORs Eq. (\ref{eq:onsager}).

Notably, in presence of a magnetic field $\boldsymbol{\mathcal B}$, the time-reversal symmetry generally breaks, meaning that generally the ORs, Eq. (\ref{eq:onsager}) are not valid in presence of a magnetic field. However, generally, the less stringent ``extended time reversal symmetry'' survives
\begin{align}
H(\boldsymbol{\mathcal B})\Theta = \Theta H(-\boldsymbol{\mathcal B})\, ,
\label{eq:microrev-extended}
\end{align}
where we explicitly expressed the dependence of $H$ on $\boldsymbol{\mathcal B}$. The latter implies the following ``extended'' microreversibility principle
\begin{align}
U_{-t}(\boldsymbol{\mathcal B}) = \Theta^\dagger U_{t}(\mathbf{-B}) \Theta\, ,
\label{eq:microrev}
\end{align}
where $U_{t}(\boldsymbol{\mathcal B})= e^{-i H(\boldsymbol{\mathcal B})t/\hbar}$. It means that in order to reverse the dynamics one needs not only to reverse the momenta and spins, but also the external field. The extended microreversibility then ensures the validity of the celebrated Onsager-Casimir relations (OCRs) \cite{Casimir45RMP17,Kubo57aJPSJ12,Bochkov79SPJETP49,Bochkov77SPJETP45,Andrieux08PRL100,deGroot84Book}:
\begin{align}
L_{A B}(t,\boldsymbol{\mathcal B}) &= \vartheta_A \vartheta_B L_{B A} (t,-\boldsymbol{\mathcal B})\, ,
\label{eq:onsager-casimir}
\end{align}
that link relaxation functions taken at opposite values of $\boldsymbol{\mathcal B}$. Here the argument $\boldsymbol{\mathcal B}$ is added to the function $L_{AB}$ to denote that it refers to the time evolution, $U_t(\boldsymbol{\mathcal B})$, relative to $H(\boldsymbol{\mathcal B})$.

Recent research has highlighted the interesting fact that the dynamical evolution of a system (classical or quantum) can be reversed in a number of different ways that differ from the application of $\Theta$, or, in case there is a magnetic field, the joint reversal of $\boldsymbol{\mathcal B}$ and the application of $\Theta$  \cite{Bonella14EPL108,Gregorio17SYM9,Coretti18MP116,Coretti20PRE102,Luo20PRR2,Carbone22AP441}. Most notably, such transformations may well not involve the reversal of the magnetic field.

For quantum systems, it is in fact straightforward to note that if the Hamiltonian, $H(\boldsymbol{\mathcal B})$, is invariant under the action of a generic antiunitary operator $K$, i.e., if there exists an antiunitary $K$ such that 
\begin{align}
H(\boldsymbol{\mathcal B}) K = K H(\boldsymbol{\mathcal B}) \, ,
\label{eq:HK=KH}
\end{align}
then 
\begin{align}
U_{-t}(\boldsymbol{\mathcal B}) =K U_{t}(\boldsymbol{\mathcal B}) K^\dagger \, ,
\label{eq:fixedB-microrev}
\end{align}
meaning that any anti-unitary symmetry of the Hamiltonian realises an inversion of the time evolution.  
We shall refer to this as an ``unconventional time reversal''. A direct aftermath of an ``unconventional time reversal'' symmetry is a set of  ``false Onsager relations" (FORs)
\begin{align}
L_{A B}(t,\boldsymbol{\mathcal B}) &= \kappa_A \kappa_B L_{B A} (t,\boldsymbol{\mathcal B})\, ,
\label{eq:constant-B-Onsager}
\end{align}
where $A,B$ are operators with well defined parity under $K$, that is
\begin{align}
K^\dagger A K = \kappa_A A,\quad K^\dagger B K = \kappa_B B\, .
\end{align}
The proof follows exactly the same standard proof of the OR's, Eq. (\ref{eq:onsager}) with $\Theta$ being replaced by $K$. The relations in Eq. (\ref{eq:constant-B-Onsager}) were first discoverd in Ref. \cite{Bonella14EPL108} for classical systems, and then reported in Ref. \cite{Gregorio17SYM9,Carbone22AP441} for special classes of quantum systems. 

At first sight one might confuse the FOR's, Eq. (\ref{eq:constant-B-Onsager}), for the OR's, Eq. (\ref{eq:onsager}), in magnetic field. A closer scrutiny however reveals that the two sets of relations substantially differ one from the other in that they refer to quantities $A,B$ that have different symmetries, a fact that was typically overlooked so far in the literature.
Generally, given two quantities $A$ and $B$ that have definite parity under $\Theta$ they might not have definite parity under $K$, or viceversa. In those cases the validity of the FORs, Eq. (\ref{eq:constant-B-Onsager}), does say nothing about the validity of the OR's, Eq. (\ref{eq:onsager}).
Viceversa, if the Hamiltonian has the $K$ symmetry, Eq. (\ref{eq:HK=KH}), two observables $A$ and $B$ have definite parity under both $\Theta$ and $K$, and $\vartheta_A \vartheta_B=\kappa_A \kappa_B$, then we have 
\begin{align}
L_{A B}(t,\boldsymbol{\mathcal B}) &= \kappa_A \kappa_B L_{B A} (t,\boldsymbol{\mathcal B}) = \vartheta_A \vartheta_B (t,\boldsymbol{\mathcal B})\, ,
\end{align}
namely, the ORs Eq. (\ref{eq:onsager}) would be valid despite the Hamiltonian does not have the $\Theta$ symmetry. However, that occurrence would only be accidental as the ORs, Eq. (\ref{eq:onsager}), would generally be violated if only one looks at different quantities $A,B$: hence our expression ``false Onsager relations''.

We shall illustrate all this below with explicit examples of interacting spin systems in magnetic field. 

\section{Omnipresence of false Onsager relations}

For quantum systems, the relations in Eq. (\ref{eq:constant-B-Onsager}) were first reported in Ref. \cite{Gregorio17SYM9} for interacting spin-less particles in homogeneous magnetic field, while Ref. \cite{Carbone22AP441} reported them for non-interacting particles with spin. 
Here we remark a crucial fact, namely that at least one ``unconventional time reversal symmetry'' can always be found in any quantum system.
To see that, recall that any antiunitary operator can be expressed as the product of the complex conjugation operator $\mathcal{K}_\mathcal{R}$ relative to some representation $\mathcal{R}$, and a generic unitary $V$ \cite{Messiah62Book}. Since any unitary can be understood as a change of basis operator, then any antiunitary operator can be understood as the complex conjugation operation relative to a specific representation.
Thus, if there exist a representation $\mathcal{R}$ in which the Hamiltonian is real, then the FORs, Eq. (\ref{eq:constant-B-Onsager}), would hold for operators having definite parity under the complex conjugation, $\mathcal{K}_\mathcal{R}$, relative to that representation.
Note that the Hamiltonian is a quantum observable, namely a Hermitian operator with real eigenvalues. Accordingly, in the representation where $H(\boldsymbol{\mathcal B})$ is diagonal, the Hamiltonian is trivially real. That is, irrespective of the specific form of the Hamiltonian, the following always holds
\begin{align}
H(\boldsymbol{\mathcal B}) = K_{H(\boldsymbol{\mathcal B})}^\dagger H(\boldsymbol{\mathcal B}) K_{H(\boldsymbol{\mathcal B})}\, ,
\label{eq:trivial-false-time-reversal}
\end{align}
where $K_{H(\boldsymbol{\mathcal B})}$ is the complex conjugation relative to the representation where $H(\boldsymbol{\mathcal B})$ is diagonal.
 It follows that for any couple of operators $A$ and $B$ that are either purely real or purely imaginary in that representation, the FORs, Eq. (\ref{eq:constant-B-Onsager}), hold. For a sufficiently complex Hamiltonian this may well have no consequence whatsoever in regard to the validity of the ORs, Eq. (\ref{eq:onsager}), which refers to quantities $A,B$ with definite parity under $\Theta$. This is further  illustrated below.

\section{Examples}
Consider an Heisenberg magnet in a possibly non homogeneous field $\boldsymbol{\mathcal B}_i=({\mathcal B}_i^x,{\mathcal B}_i^y,{\mathcal B}_i^z)$,
\begin{align}
H(\boldsymbol{\mathcal B}) = J \sum_{i,j} \boldsymbol{\sigma}_i \cdot \boldsymbol{\sigma}_j - \sum_i \boldsymbol{\mathcal B}_i \cdot \boldsymbol{\sigma}_i\, ,
\label{eq:HeisenbergH}
\end{align}
where $\boldsymbol{\sigma}_i= (\sigma_i^x,\sigma_i^y,\sigma_i^z)$, with $\sigma_i^\alpha$ denoting Pauli operators.
Since, by definition, all spin operators $\sigma_i^\alpha$ are odd under time reversal $\Theta$, Eq. (\ref{eq:timerev}),
the Hamiltonian is not invariant under time reversal $\Theta$, while it is invariant under the joint action of $\Theta$, and the reversal of $\boldsymbol{\mathcal B}$, i.e., it obeys Eq. (\ref{eq:microrev-extended}). Accordingly the OCRs, Eq. (\ref{eq:constant-B-Onsager}) are obeyed. 

Let us consider first the homogeneous case $\boldsymbol{\mathcal B}_i=\boldsymbol{\mathcal B}$. Let's fix the axes so that $z$ is the direction of the applied field $\boldsymbol{\mathcal B}$, $\boldsymbol{\mathcal B}={\mathcal B}_z\hat{\mathbf{z}}$, then:
\begin{align}
H(\boldsymbol{\mathcal B}) = J \sum \boldsymbol{\sigma}_i \cdot\boldsymbol{\sigma}_j - {\mathcal B}_z \sum_i \sigma_i^z
\end{align}

In the representation where the tensor product $\otimes_i  \sigma_i^z$ is diagonal, the Hamiltonian is real. That is if  $K_{\otimes \sigma_i^z}$ is the complex conjugation relative to that representation, it is
\begin{align}
H(\boldsymbol{\mathcal B}) = K_{\otimes \sigma_i^z}^\dagger  H(\boldsymbol{\mathcal B}) K_{\otimes \sigma_i^z}\, .
\label{eq:H=KHK-sigmaz}
\end{align}
In fact, in said representation, the Pauli matrices $ \sigma_i^z$  and $\sigma^x_i$ are real
while the $\sigma^y_i$ are imaginary. Thus the term that couples to the applied magnetic field is real. The interaction term is also real because it is the sum of terms of the type $\sigma_i^\alpha \sigma_j^\alpha$, which are real regardless of whether $\alpha$ is $z,x$, or $y$.

Consider now the case when the external field is not homogeneous in space $\boldsymbol{\mathcal B}_i\neq \boldsymbol{\mathcal B}_j$ for $i \neq j$. 
If it only changes in modulo, but it has a fixed direction, then the previous argument will continue to apply unaltered.  
If its direction also changes, but remains confined onto one fixed plane, the argument still applies, with minimal changes. To see that
fix the axes so that the field has only components along $x$ and $z$, so that the Hamiltonian reads
\begin{align}
H(\boldsymbol{\mathcal B}) = J \sum \boldsymbol{\sigma}_i \cdot\boldsymbol{\sigma}_j - \sum (\mathcal{B}^x_i\sigma_i^x+\mathcal{B}^z_i\sigma_i^z)\, .
\end{align}
Since in the representation of the basis spanned by the eigenvectors of  $\otimes \sigma_i^z$ both the $\sigma_i^x$ and the $\sigma_i^z$ are all real, and, as discussed above the interaction term is also real, then the Hamiltonian is real as well in said representation, that is, the unconventional time reversal symmetry of Eq. (\ref{eq:H=KHK-sigmaz}) holds.

\begin{figure}[t]
    \centering
    \includegraphics[width=\linewidth]{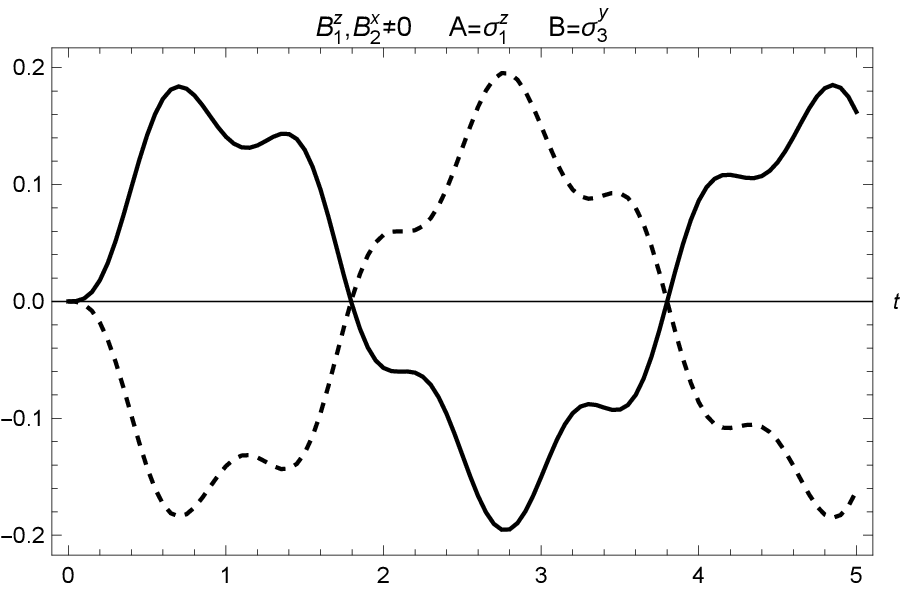}
    \includegraphics[width=\linewidth]{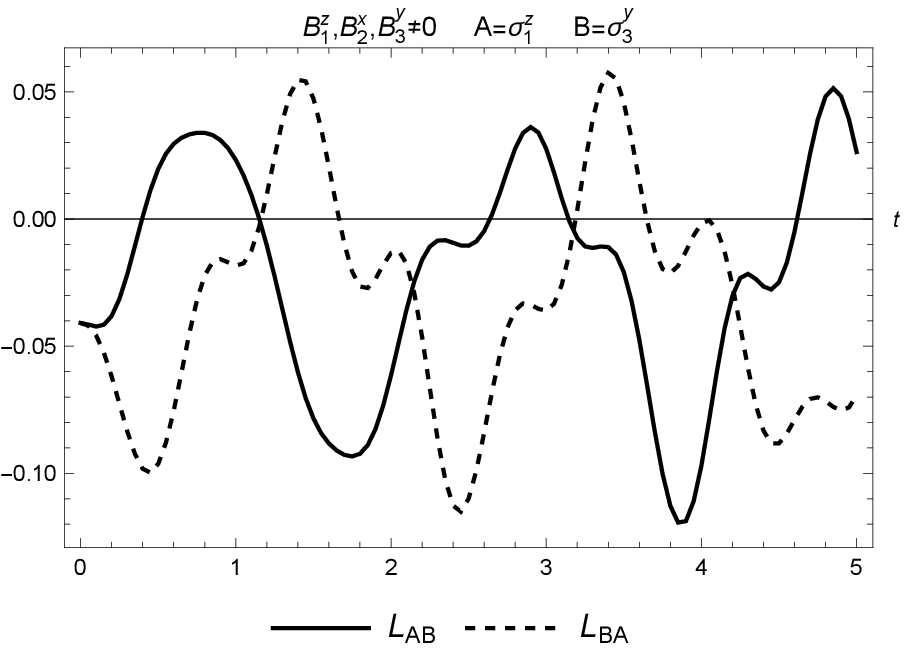}
    \caption{Breakdown of Onsager relations, Eq. (\ref{eq:onsager}). Top: Relaxation functions $L_{A B}$ and $L_{BA}$ for $A=\sigma_1^z$, $B=\sigma_3^z$ for a 3 sites Heisenberg magnet, Eq. (\ref{eq:HeisenbergH}), with ${\mathcal B}_i^\alpha=0$ except $B_1^z, B_2^x = -2$. According to Eq. (\ref{eq:Lxy=-Lyx}) it is $L_{A B}=-L_{BA}$, whereas the OR's predicts $L_{AB}=L_{BA}$. Bottom: same as top except for $B_3^y$, which now has the value $B_3^y=-2$. The ORs, Eq. (\ref{eq:HeisenbergH}), predicting $L_{AB}=L_{BA}$ is not obeyed. In both plots it is $J=1$, $\beta=2$, and $\hbar=1$.}
    \label{fig:Fig}
\end{figure}

Accorodingly, the FORs, Eq. (\ref{eq:constant-B-Onsager}), hold for any couple of observables $A,B$ that have definite parity under the transformation $K_{\otimes \sigma_i^z}$.
The spin operators $\sigma_i^\alpha$ all have definite parities under $K_{\otimes \sigma_i^z}$: the $\sigma_i^\alpha$ are even for $\alpha=x,z$, and are odd for $\alpha=y$. Considering that all the operators $\sigma_i^\alpha$ are, by definition, odd under the time reversal $\Theta$, it follows, for example, that the ORs, Eq. (\ref{eq:onsager}) would hold despite the presence of the magnetic field, for couples of operators $A= \sigma_i^\alpha$, $B= \sigma_j^\beta$, with $\alpha$ and $\beta$ being both $y$, or both either $x$ or $z$.
However for couples of the type $\alpha=y$, $\beta=x,z$, the FOR, Eq. (\ref{eq:constant-B-Onsager}), reads
\begin{align}
L_{\sigma_i^y, \sigma_j^{z(x)}}(t,\boldsymbol{\mathcal B}) = -L_{\sigma_j^{z(x)},\sigma_i^y, }(t,\boldsymbol{\mathcal B})\, ,
\label{eq:Lxy=-Lyx}
\end{align}
thus violating the ORs Eq. (\ref{eq:onsager}), which would predict a plus sign on the r.h.s. instead of a minus sign. This is illustrateed in Fig. \ref{fig:Fig}, top panel. Based on Eq. (\ref{eq:Lxy=-Lyx}) it is not hard to get convinced that the ORs, Eq. (\ref{eq:onsager}), would not hold for $A$ and $B$ being spin operators (which are odd under $\Theta$) pointing in generic directions, i.e. for  $A=\mathbf a_i \cdot \boldsymbol{\sigma}_i, B=\mathbf{b}_i \cdot \boldsymbol{\sigma}_i$ where $\mathbf{a}_i, \mathbf{b}_i$ denote real unit vectors.

Consider now the general case where $\boldsymbol{\mathcal B}_i$ is not confined to a plane but explores all spatial directions as the spatial index $i$ is varied over all sites of the spin network. We note that, at variance with the case above, in this case it is generally not even possible to find an ``unconventional time reversal'' symmetry besides the trivial invariance under $K_{H(\mathbf B)}$. So while in the previous case the OR's  Eq. (\ref{eq:onsager}) remain valid for certain couples of observables, now we expect them to be generally violated. Fig. \ref{fig:Fig}, bottom panel, shows an example of such violations.

\section{Fixed field fluctuation relations}

The presence of an ``unconventional time reversal'' symmetry may as well result in the validity of the nonequilibrium flutuation relations \cite{Campisi11RMP83} despite the presence of a magnetic field. At variance with linear response relations, those relations apply arbitrarily far from equilibrium and refer to situations where the Hamiltonian has an explicit time-dependence, $H=H(t,\boldsymbol{\mathcal B})$. Under the provision that there exist an ``unconventional time reversal'' transformation, $K$ , that commutes with the Hamiltonian at all times during the driving protocol, i.e.,
\begin{align}
K H(t,\boldsymbol{\mathcal B}) = H(t,\boldsymbol{\mathcal B}) K, \quad  \forall t \in [0,\tau],
\label{eq:KH(t)=H(t)K}\
\end{align}
one would have, e.g., for a system prepared in an equilibrium thermal state that evolves unitarily, the following ``fixed field fluctuation relation'' for work
\begin{align}
\frac{p(w,\boldsymbol{\mathcal B})}{\tilde p(-w,\boldsymbol{\mathcal B})} = e^{-\beta (w-\Delta F)},
\label{eq:falseFR}
\end{align}
featuring the same $\boldsymbol{\mathcal B}$ in both the forward work statistics, $p(w,\boldsymbol{\mathcal B})$ (obtained from the evolution generated by $H(t,\boldsymbol{\mathcal B})$), and the backward work statistics, $\tilde p(w,\boldsymbol{\mathcal B})$ (obtained from the evolution generated by $H(\tau-t,\boldsymbol{\mathcal B})$, with $\tau$ the duration of the time dependent driving). This is at variance with the case customarily discussed in the literature whereby Eq. (\ref{eq:microrev-extended}) is assumed to hold at all times \cite{Jarzynski04PRL92,Andrieux08PRL100,Campisi11RMP83}) and the backward probability $\tilde p$ is taken at reversed field  $-\boldsymbol{\mathcal B}$ .

The proof of Eq. (\ref{eq:falseFR}) follows exactly the standard proof (see appendices B, and C of Ref. \cite{Campisi11RMP83}) with $\Theta$ being replaced by $K$. Note that at variance with the Onsager relations, that are possibly only apparently obeyed in case of ``false time reversal violation'', the fluctuation relation $p(w)/p(-w)=e^{-\beta (w-\Delta F)}$ is simply obeyed in its standard form, despite the magnetic field, when Eq. (\ref{eq:KH(t)=H(t)K}) holds.

The validity of Eq. (\ref{eq:falseFR}) was often observed in the literature, see e.g., \cite{Dorner12PRL109,Fusco14PRX4,Schmidtke18PRE98}, although the issue related to the presence of a magnetic field and the according lack of microreversibility (which would have ensured its validity) typically passed unnoticed. Needless to say, all those previous works considered situations where the Hamiltonian was real in some representation, at all times, which ensured the validity  of Eq. (\ref{eq:falseFR}), a fact that instead was only acknowledged and discussed in Ref. \cite{Schmidtke18PRE98}.

In this regard it is also worth remarking that Eq. (\ref{eq:fixedB-microrev}) would instantaneously hold at each time, $t$, with an ``unconventional time reversal'' $K_{H(t,\boldsymbol{\mathcal B})}$, that possibly changes in time. Accordingly, its ubiquitous validity does not ensure the validity of Eq. (\ref{eq:falseFR}) featuring a fixed $K$, Eq. (\ref{eq:KH(t)=H(t)K}). It is a simple exercise to show that, even for a single spin in a time dependent magnetic field $H(t,\boldsymbol{\mathcal B})=-\boldsymbol{\mathcal B}(t) \cdot \boldsymbol{\sigma}$, Eq. (\ref{eq:falseFR}) does not hold when $\boldsymbol{\mathcal B}(t)$ explores all three spatial dimensions
\cite{Orlandini22UNIFI}.

\section{Discussion}
We have established two main facts. 
First,  for quantum systems a trivial ``unconventional time reversal'' always exist, and that is the symmetry under complex conjugation in the representation where the Hamiltonian is diagonal. This very same fact was already noticed by Robnik and Berry \cite{Robnik86JPA19} in the context of level statistics in quantum billiards. 
Second, in contrast with what previous research suggested, the presence of one or more ``unconventional time reversals'' does not by itself guarantee the validity of the Onsager relations, Eq. (\ref{eq:onsager}), when a magnetic field is present.  As illustrated by examples, they indeed can be violated, however if couples of observables exist that have definite parities under both time reversal and an ``unconventional time reversal'' which is a symmetry of the problem, the ORs will be obeyed for them, provided the products of their parities under time reversal is the same as that under the ``unconventional time reversal''. 

Roughly speaking, the more ``unconventional time reversal symmetries'' a system has, the larger the set of couples of observables that obey the ORs, Eq. (\ref{eq:onsager}), despite a magnetic field. On the contrary when the only ``unconventional time reversal'' is the trivial one, Eq. (\ref{eq:trivial-false-time-reversal}) one should generally expect the ORs Eq. (\ref{eq:onsager}) to be violated. The situation is somewhat similar to dynamical system theory, where we have two extremes: full integrability (as many conserved quantities as are the degrees of freedom), and full ergodicity (the Hamiltonian is the only first integral of motion). Between these two extremes lie complex systems displaying both regular and irregular motion. Here, similarly, we have the case where there are as many ``unconventional time reversal symmetries" as are the degrees of freedom \footnote{This would occur, for example, in the case of a set of non-interacting spins with local magnetic field. In that case all relaxation functions are null and the ORs are trivially obeyed for any couple of observables.
}, the case when only the trivial one exist, and the complex situation in between featuring both violation and obedience of the ORs, Eq. (\ref{eq:onsager}).

Since, as illustrated by our examples, the Onsager relations may be violated despite the presence of ``false time reversal violations'', our results leave open  the possibility of the ``Umkehreffekt'', which in fact, as mentioned above, has been experimentally observed. Thus the present study does not provide any fundamental reason to exclude the possibility of achieving Carnot efficiency at finite power, in the way discussed in Ref. \cite{Benenti11PRL106}.

We further have commented that the omnipresence of ``false time reversal violations'' does not, per se, imply the ubiquitous validity of quantum fluctuation relations, e.g., the work fluctuation relation, in presence of a magnetic field. The latter would hold provided the ``unconventional time reversal symmetry'' is one and the same during the whole driving protocol. While many models of many-body systems that are customarily studied 
in the literature (e.g., the driven Ising model in transverse field)  satisfy that requirement that is generally not the case.

A question that remains to be answered is whether ``false time reversal violation'' is omnipresent as well in the classical case, and how can one construct the according ``unconventional time reversal'' transformation. While this issue was easily addressed in the quantum case, the question does not seem to admit a simple answer for classical systems. Addressing that might reveal a new discordance between classical and quantum realms \footnote{One such discordance whereby deterministic friction appears in the classical realm but not in the quantum one was reported by Berry and Robbins \cite{Berry93PRSLA442}}.

\section{Acknowledgements}
I would like to thank Giuliano Benenti, Lamberto Rondoni and Davide Carbone  for pointing out relevant literature and for the stimulating discussions that were instrumental to shaping this work.

\bibliographystyle{eplbib}


\end{document}